\documentclass[aps,prd,superscriptaddress,twoside,twocolumn,nofootinbib,showpacs,floatfix]{revtex4-1}
\usepackage{amsmath,amssymb}
\usepackage{graphicx}
\usepackage{subfigure}
\usepackage{braket}
\usepackage{slashed}
\usepackage[usenames,dvipsnames]{color}
\usepackage[colorlinks]{hyperref}
\allowdisplaybreaks

\begin{document}

\title{Exploring $K\Xi^*$ and $K^*\Xi$ molecular states and the triangle singularity in the $K^- p \to K \Xi(1530)$ reaction}

\author{Ke Wang}
\email{wangke1@mail.tsinghua.edu.cn}
\affiliation{Department of Physics and Center for High Energy Physics, Tsinghua University, Beijing 100084, China}

\author{Fei Huang}
\email{huangfei@ucas.ac.cn}
\affiliation{School of Nuclear Science and Technology, University of Chinese Academy of Sciences, Beijing 101408, China}

\author{Bing-Song Zou} 
\email{zoubs@mail.tsinghua.edu.cn}
\affiliation{Department of Physics and Center for High Energy Physics, Tsinghua University, Beijing 100084, China}
%\affiliation{CAS Key Laboratory of Theoretical Physics, Institute of Theoretical Physics, Chinese Academy of Sciences, Beijing 100190, China}

\date{\today}

\begin{abstract}
We investigate the $K^- p \to K \Xi(1530)$ reaction within an effective Lagrangian approach, exploring possible $K \Xi^*$ and $K^* \Xi$ hadronic molecular states and the role of the triangle singularity (TS). The $\Lambda(2050)3/2^-$ is interpreted as a $K \Xi^*$ molecule, whereas a $K^* \Xi$ molecule with $I(J^P)=0(3/2^-)$ and mass about 2150~MeV denoted as $\Lambda(2150)$ can generate a TS through triangle-loop diagrams with intermediate $K^*$, $\Xi$, and $\pi$. The peak structure observed in the cross section near $\sqrt{s}=2.25$~GeV is analyzed in terms of both the $\Sigma(2250)$ resonance production and the TS mechanism associated with $\Lambda(2150)$. We find that the TS induces pronounced spin effects in the final state $\Xi^*$, which can be probed through measurements of its spin density matrix elements. In particular, significant variations of the spin observables in the $\sqrt{s}=2.2$--$2.3$~GeV region serve as a distinct TS signature absent in a pure resonance scenario. Furthermore, for the three-body reaction $K^- p \to K^+ \pi^- \Xi^0$, we demonstrate that $\Xi^*$ spin observables can be reliably extracted from the $\pi$ angular distribution in the $\Xi \pi$ rest frame by applying an appropriate kinematic cut on the $\Xi\pi$ invariant mass to suppress background contributions. These predictions can be tested in future high-precision measurements at J-PARC, providing crucial insights into the nature of the TS and the possible existence of the $K^* \Xi$ molecular state.

\end{abstract}

%\pacs{xxx,xxx,xxx,xxx}

\maketitle

\section{Introduction}   \label{sec:int}

Hadronic molecules represent an important frontier in hadron physics. In the conventional quark model~\cite{Gell-Mann:1964ewy,Zweig:1964ruk}, hadrons are classified as mesons with each composed of a quark-antiquark pair and baryons with each composed of three quarks. Extending beyond this framework, hadronic molecules feature multiquark configurations ascribed as loosely bound systems of two conventional hadrons, held together by residual strong interactions mediated by hadron exchange. Such states typically lie near the corresponding two-hadron thresholds, in close analogy to the deuteron. With growing numbers of near-threshold structures identified as candidates for hadronic molecules~\cite{Chen:2016qju,Guo:2017jvc,Dong:2021juy,Meng:2022ozq}, establishing experimental observables that are sensitive to their internal composition is essential for clarifying the nature of these exotic states.

Meson-baryon molecular candidates in the heavy-quark sector have been widely investigated~\cite{Wu:2010jy,Wu:2010vk,Wang:2011rga,Wu:2012md,Olsen:2017bmm,Ali:2017jda,Garcilazo:2025wkt,Husken:2024rdk}. In 2015, the LHCb Collaboration observed the $P_c(4380)$ and $P_c(4450)$ structures, with a minimum quark content of $c\bar{c}uud$, in the $J/\psi p$ invariant mass spectrum of $\Lambda^0_b \to J/\psi K^- p$ decays~\cite{LHCb:2015yax}. A subsequent analysis of a larger data sample in 2019~\cite{LHCb:2019kea} led to the discovery of the $P_c(4312)$ and resolved the previously observed $P_c(4450)$ into two narrow resonances, $P_c(4440)$ and $P_c(4457)$, while the status of the $P_c(4380)$ remains inconclusive. The proximity of these $P_c$ states to the $\bar{D}^{(*)} \Sigma_c^{(*)}$ thresholds strongly favors their interpretation as molecular bound states of $\bar{D}^{(*)}$ and $\Sigma_c^{(*)}$ hadrons. This picture is further supported by extensive theoretical studies from both hadronic-level~\cite{Chen:2019asm,He:2019ify,He:2019rva,Dong:2021bvy,Wang:2025ecf} and quark-level~\cite{Huang:2019jlf,Liu:2019tjn,Wang:2024unj} approaches.

In the light-quark sector, the $N(1875)$, $N(2080)$, and $N(2270)$ resonances have been proposed as strange partners of the $P_c$ states, interpreted as $K^{(*)} \Sigma^{(*)}$ molecular configurations arising from the replacement of the $c\bar{c}$ pair in $\bar{D}^{(*)} \Sigma_c^{(*)}$ systems with an $s\bar{s}$ pair~\cite{He:2017aps,Zou:2018uji,Lin:2018kcc,Wu:2023ywu,Ben:2023uev,Suo:2025rty,Tian:2025bkx}. Furthermore, Ref.~\cite{Dong:2021bvy} predicted isoscalar ($I=0$) hidden-charm molecular states formed from $\bar{D}^{(*)} \Xi_c^{(*)}$ interactions. By analogy, $K^{(*)} \Xi^{(*)}$ (with $\Xi^* \equiv \Xi(1530)$ hereafter) systems, the strange counterparts of $\bar{D}^{(*)} \Xi_c^{(*)}$, are expected to exhibit sufficiently attractive interactions to support molecular state formation. 

In this work, we focus on two possible $S$-wave $\Lambda^*$ molecular states: a $K \Xi^*$ bound state and a $K^* \Xi$ state. The $\Lambda(2050)3/2^-$, lying close to the $K \Xi^*$ threshold ($2029$ MeV), provides a natural candidate for the former. On the other hand, if a molecular $K^* \Xi$ state exists below the $K^* \Xi$ threshold ($2212$ MeV), pion exchange between the $K^*$ and $\Xi$ would strongly couple this state to the $K \Xi^*$ channel through triangle-loop diagrams, thereby generating a triangle singularity (TS). Therefore, both molecular states can be investigated through $K \Xi^*$ production, and the observation of a TS signal in the corresponding process would offer indirect evidence for the existence of the $K^* \Xi$ molecular state. 

The triangle singularity was first pointed out by Landau in 1959~\cite{Landau:1959fi}. Subsequent theoretical work has demonstrated its crucial role in resolving longstanding puzzles in hadron spectroscopy~\cite{Wu:2011yx,Aceti:2012dj,Wu:2012pg,Achasov:2015uua,Du:2019idk,Jing:2019cbw,Guo:2019qcn,Sakai:2020ucu,Molina:2020kyu,Sakai:2020crh,Yan:2022eiy,Wang:2022wdm,Wang:2023xua,Wang:2024ewe}. The physical picture of TS production corresponds to a classical process, as described by the Coleman-Norton theorem~\cite{Coleman:1965xm}. In the present case, it arises in the decay of a $K^* \Xi$ molecular state into $K \Xi^*$ via a triangle loop with intermediate $K^*$, $\Xi$, and $\pi$. When these three particles are simultaneously on shell and carry collinear three-momenta, the pion travels in the same direction as the $\Xi$, catches up with it, and fuses into the $\Xi^*$. This classical process produces a pronounced peak in the $K \Xi^*$ invariant mass spectrum near $2.25$~GeV. Apart from the cross-section enhancement, the restrictive kinematics of the TS can induce significant spin effects in the final state. These distinctive features can be exploited to discriminate between different reaction mechanisms and to search for possible hadronic resonances~\cite{Wang:2022wdm,Wang:2023xua,Wang:2024ewe}.

The $K^- p \to K \Xi(1530)$ reaction provides an ideal platform for investigating possible $K \Xi^*$ and $K^* \Xi$ molecular states. Although Ref.~\cite{Guo:2025ibo} attributed the two peaks observed near $\sqrt{s}=2.1$~GeV and $2.25$~GeV to resonance production, we propose an alternative scenario in which the higher-mass peak may arise from the TS mechanism associated with a $K^* \Xi$ molecular state. To this end, we introduce two phenomenological models to fit the experimental data: Model I interprets the peak at $\sqrt{s}=2.25$~GeV within a resonance-production framework, whereas Model II ascribes it to the TS mechanism. The characteristic spin effects generated by the TS in $\Xi^*$ production offer a possible means to discriminate between these two scenarios. Furthermore, because extracting these observables requires angular distribution data from the three-body process $K^- p \to K \pi \Xi$, we extend our analysis to this channel and examine how background contributions affect the extraction of spin observables. Our results demonstrate that such measurements can unambiguously determine the origin of the peak at $\sqrt{s}=2.25$~GeV, thereby providing decisive evidence either for or against the existence of a $K^* \Xi$ molecular state.

This paper is organized as follows. In Sec.~\ref{sec:th}, we present the theoretical framework and amplitudes for the reaction $K^- p \to K \Xi(1530)$. In Sec.~\ref{sec:res}, we show the numerical results and discuss their implications. Finally, we summarize our findings and conclusions in Sec.~\ref{sec:sum}.

\section{Theoretical framework}\label{sec:th}

\begin{figure}[tbp]
      \includegraphics[width=1.0\columnwidth]{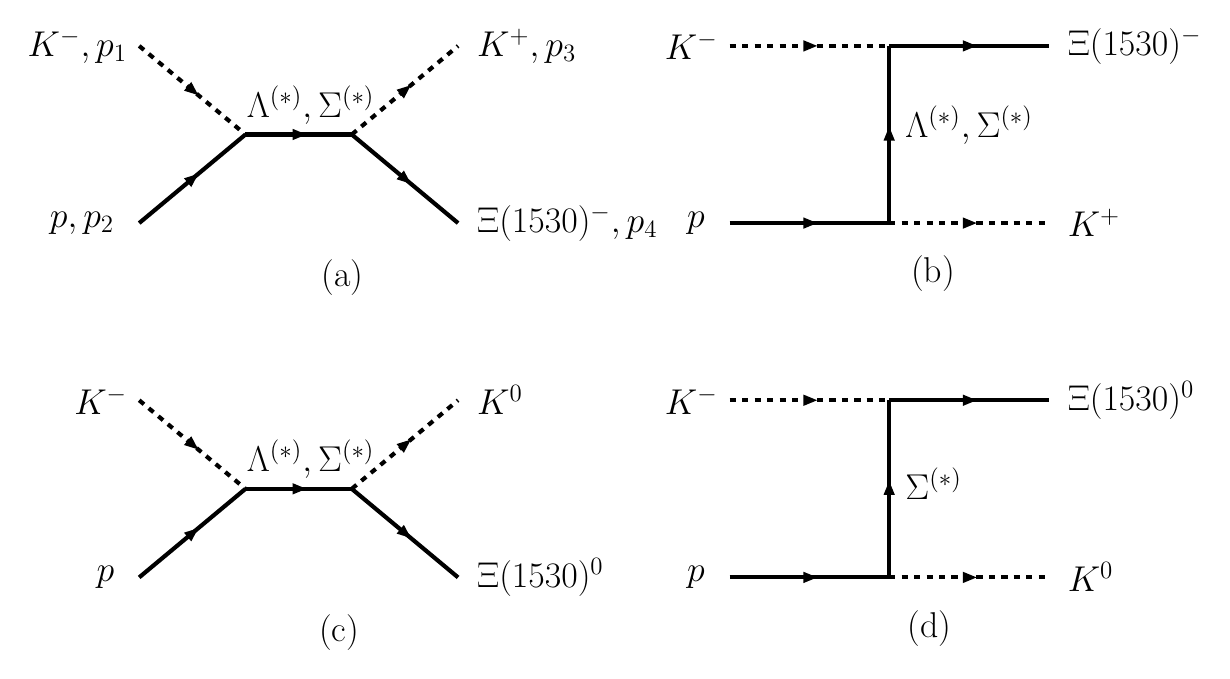}
      \caption{Tree-level Feynman diagrams for $K^- p \to K^+ \Xi(1530)^-$ and $K^- p \to K^0 \Xi(1530)^0$.}
      \label{fig:tree}
\end{figure}

\begin{figure}[tbp]
      \includegraphics[width=0.8\columnwidth]{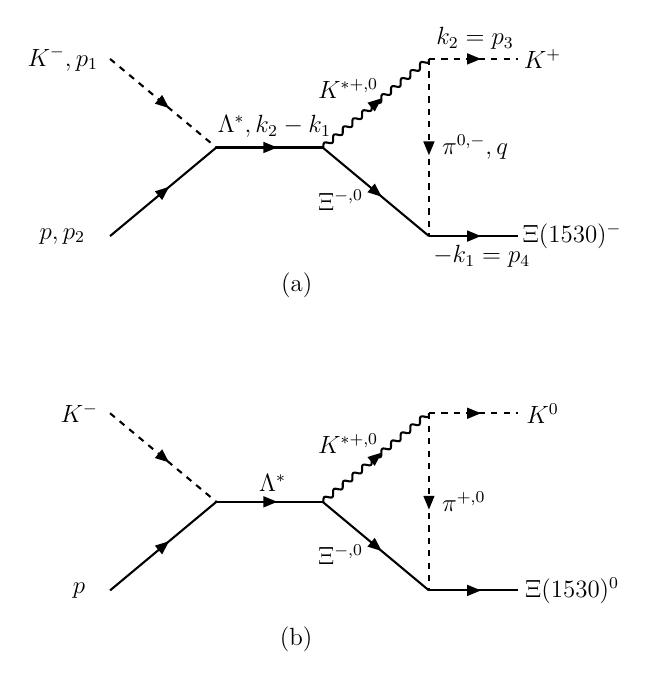}
	\caption{Triangle-loop Feynman diagrams for $K^- p \to K^+ \Xi(1530)^-$ and $K^- p \to K^0 \Xi(1530)^0$. }
	\label{fig:loop}
\end{figure}

In this work, we study the $K^- p \to K \Xi(1530)$ reaction within an effective Lagrangian approach and the isobar model. As discussed in Sec.~\ref{sec:int}, the possible $K\Xi^*$ and $K^*\Xi$ molecular states can be investigated in this reaction. Meanwhile, the peak observed near $2.25$~GeV in the total cross sections may arise from either resonance production or triangle-loop diagrams involving intermediate $K^* \Xi \pi$ states. To calculate their contributions and distinguish these mechanisms, we construct two phenomenological models: Model I contains only tree-level contributions, with the peak generated by resonance production, whereas Model II additionally incorporates triangle-loop diagrams and attributes the peak to the TS mechanism. The corresponding Feynman diagrams are displayed in Figs.~\ref{fig:tree} and \ref{fig:loop}, respectively.

For the tree-level diagrams, $t$-channel processes are forbidden because no meson carrying double strangeness exists to be exchanged. The $\Lambda$ and $\Sigma$ hyperons and their excited states play an important role in $s$- and $u$-channel processes, and all intermediate particles used in this work are listed in Table~\ref{tab:particles}. Here we take the lower-limit value of the $\Lambda(2050)$ mass from the PDG~\cite{ParticleDataGroup:2024cfk}, rather than the central value, to reflect its possible $K\Xi^*$ molecular-state nature. Its precise determination remains to be settled by future high-statistics data. The reasons for selecting these states are discussed in the following sections. Furthermore, as $u$-channel contributions from high-spin hyperon resonances are negligible above threshold~\cite{Sharov:2011xq}, we retain only the $s$-channel contributions for $\Lambda(2110)$ and $\Sigma(2250)$. 

The effective Lagrangian densities for these tree-level processes take the form~\cite{Guo:2025ibo}
\begin{eqnarray}
	\mathcal{L}_{Y_{1/2^+} N K} \!\!\! &=& \!\!\! \frac{f_{Y_{1/2^+} N K}}{m_\pi} \bar{Y}_{1/2^+} \gamma_{\mu} \gamma_{5} \partial^{\mu} \bar{K} N + \mathrm{H.c.}, \\
    \mathcal{L}_{Y_{3/2^-} N K} \!\!\! &=& \!\!\! \frac{f_{Y_{3/2^-} N K}}{m_\pi} \bar{Y}^\mu_{3/2^-} \partial_\mu K \gamma_{5} N + \mathrm{H.c.}, \\
    \mathcal{L}_{Y_{5/2^{\pm}} N K} \!\!\! &=& \!\!\! -i \frac{f_{Y_{5/2^{\pm}} N K}}{m^2_\pi} \bar{Y}^{\mu\nu}_{5/2^{\pm}} \partial_\mu \partial_\nu K \left\{\begin{array}{c} \gamma_{5} \\ 1 \end{array}\right\} N \nonumber \\ & & \!\!\! + \, \mathrm{H.c.}, \\
    \mathcal{L}_{Y_{1/2^+} \Xi^* K} \!\!\! &=& \!\!\! \frac{f_{Y_{1/2^+} \Xi^* K}}{m_\pi} \bar{Y}_{1/2^+} \Xi^*_{\mu} \partial^{\mu} K + \mathrm{H.c.}, \\
    \mathcal{L}_{Y_{3/2^-} \Xi^* K} \!\!\! &=& \!\!\! i f_{Y_{3/2^-} \Xi^* K} \bar{Y}^\mu_{3/2^-} \Xi^{*}_\mu K + \mathrm{H.c.}, \\
    \mathcal{L}_{Y_{5/2^{\pm}} \Xi^* K} \!\!\! &=& \!\!\! \frac{f_{Y_{5/2^{\pm}} \Xi^* K}}{m_\pi} \bar{Y}^{\mu\nu}_{5/2^{\pm}} \left\{\begin{array}{c} 1 \\ \gamma_{5} \end{array}\right\} \Xi^*_{\mu} \partial_\nu K \! + \! \mathrm{H.c.},
\end{eqnarray}
where $Y_{J^P}$ denotes a hyperon state with spin-parity $J^P$, and the upper and lower symbols in the curly brackets correspond to the positive and negative parity cases, respectively.

\begin{table}[htb]
      \caption{The hyperons adopted in this work. Experimental masses and decay widths are taken from the PDG~\cite{ParticleDataGroup:2024cfk}. }
      \renewcommand{\arraystretch}{1.2}
      \begin{tabular*}{\columnwidth}{@{\extracolsep\fill}ccccc}
      \hline\hline
      States & $J^P$ & Mass (MeV) & Width (MeV)\\
      \hline
      $\Lambda(1116)$ & $1/2^+$ & 1116 & ...\\ 
      $\Lambda(2050)$ & $3/2^-$ & 2034 & 550\\ 
      $\Lambda(2110)$ & $5/2^+$ & 2090 & 250\\
      $\Lambda(2325)$ & $3/2^-$ & 2325 & 177\\ 
      $\Sigma(1193)$ & $1/2^+$ & 1193 & ...\\ 
      $\Sigma(2250)$ & $5/2^-$ & 2250 & 100\\ 
      \hline\hline
      \end{tabular*}
      \label{tab:particles}
\end{table}

To account for the internal structure of the interaction vertices and the off-shell effects of intermediate particles, we introduce phenomenological form factors. For intermediate mesons and baryons, these are parametrized as
\begin{eqnarray}
	F_{M}\left(q,m\right) &=& \frac{\Lambda^2_M - m^2}{\Lambda^2_M - q^2}, \\
	F_{B}\left(q,m\right) &=& \frac{\Lambda^4_B}{\Lambda^4_B + \left( q^2 - m^2 \right)^2},
\end{eqnarray}
where $q$ and $m$ are the four-momentum and mass of the exchanged particle, respectively. To keep the number of free parameters minimal, we fix the cutoff value $\Lambda_B = 0.5$~GeV for high-spin intermediate states with $J \ge 5/2$~\cite{Sharov:2011xq,Guo:2025ibo}, while the cutoff for the remaining low-spin states is treated as a free parameter. The propagators for the exchanged particles of spin $J$ take the form
\begin{eqnarray}
	G_0(q) &=& \frac{i}{q^2-m^2}, \\
	G_{1/2}(q) &=& \frac{i(\slashed{q}+m)}{q^2-m^2+im\Gamma}, \\
	G^{\mu\nu}_{1}(q) &=& -\frac{i \bar{g}^{\mu\nu}}{q^2-m^2+im\Gamma}, \\
	G^{\mu\nu}_{3/2}(q) &=& \frac{i(\slashed{q}+m)}{q^2-m^2 +im\Gamma} \left[-g^{\mu\nu}+\frac{1}{3}\gamma^\mu\gamma^\nu \right. \nonumber \\ & & \left.+ \, \frac{1}{3m}(\gamma^\mu q^\nu- \gamma^\nu q^\mu)+\frac{2}{3m^2}q^\mu q^\nu\right], \\
	G^{\mu\nu\alpha\beta}_{5/2}(q) &=& \frac{i(\slashed{q}+m)}{q^2-m^2 +im\Gamma} \Bigg[\frac{1}{2} \left( \bar{g}^{\mu\alpha} \bar{g}^{\nu\beta} + \bar{g}^{\mu\beta} \bar{g}^{\nu\alpha} \right) \nonumber \\ & &  - \, \frac{1}{5} \bar{g}^{\mu\nu} \bar{g}^{\alpha\beta} - \frac{1}{10} ( \bar{\gamma}^\mu \bar{\gamma}^\alpha \bar{g}^{\nu\beta} + \bar{\gamma}^\mu \bar{\gamma}^\beta \bar{g}^{\nu\alpha} \nonumber \\ & & + \, \bar{\gamma}^\nu \bar{\gamma}^\alpha \bar{g}^{\mu\beta} + \bar{\gamma}^\nu \bar{\gamma}^\beta \bar{g}^{\mu\alpha} ) \Bigg],
\end{eqnarray}
where the notations $\bar{g}^{\mu\nu}$ and $\bar{\gamma}^\mu$ are defined as in Ref.~\cite{Kim:2017nxg}:
\begin{equation}
	\bar{g}^{\mu\nu} = g^{\mu\nu} - \frac{q^\mu q^\nu}{m^2}, \quad
	\bar{\gamma}^\mu = \gamma_\mu - \frac{q^\mu}{m^2} \slashed{q}.
\end{equation}

Using the ingredients introduced above, we construct the tree-level amplitudes for the diagrams shown in Fig.~\ref{fig:tree} following the standard procedure. We obtain
\begin{eqnarray}
	\mathcal{M}^s_{Y_{1/2^+}} \!\! &=& f_I \frac{f_{Y_{1/2^+} N K} f_{Y_{1/2^+} \Xi^* K}}{m^2_\pi} \bar{u}_{\Xi^*,\mu} p^\mu_3 G_{1/2}(p_s) \nonumber \\ & & \times \, \slashed{p}_1 \gamma_{5} u_p F(p_s), 
	\\
	\mathcal{M}^u_{Y_{1/2^+}} \!\! &=& f_I \frac{f_{Y_{1/2^+} N K} f_{Y_{1/2^+} \Xi^* K}}{m^2_\pi} \bar{u}_{\Xi^*,\mu} p^\mu_1 G_{1/2}(p_u) \nonumber \\ & & \times \, \slashed{p}_3 \gamma_{5} u_p F(p_u), 
	\\
	\mathcal{M}^s_{Y_{3/2^-}} \!\! &=& f_I \frac{f_{Y_{3/2^-} N K} f_{Y_{3/2^-} \Xi^* K}}{m_\pi} \bar{u}_{\Xi^*,\mu} G^{\mu\nu}_{3/2}(p_s) \nonumber \\ & & \times \, p_{1,\nu} \gamma_{5} u_p F(p_s), 
	\\
	\mathcal{M}^u_{Y_{3/2^-}} \!\! &=& -f_I \frac{f_{Y_{3/2^-} N K} f_{Y_{3/2^-} \Xi^* K}}{m_\pi} \bar{u}_{\Xi^*,\mu} G^{\mu\nu}_{3/2}(p_u) \nonumber \\ & & \times \, p_{3,\nu} \gamma_{5} u_p F(p_u), 
	\\
	\mathcal{M}^s_{Y_{5/2^+}} \!\! &=& -f_I \frac{f_{Y_{5/2^+} N K} f_{Y_{5/2^+} \Xi^* K}}{m^3_\pi} \bar{u}_{\Xi^*,\mu} p_{3,\nu} G^{\mu\nu\alpha\beta}_{5/2}(p_s) \nonumber \\ & & \times \, p_{1,\alpha} p_{1,\beta} \gamma_{5} u_p F(p_s), 
	\\
	\mathcal{M}^u_{Y_{5/2^+}} \!\! &=& f_I \frac{f_{Y_{5/2^+} N K} f_{Y_{5/2^+} \Xi^* K}}{m^3_\pi} \bar{u}_{\Xi^*,\mu} p_{1,\nu} G^{\mu\nu\alpha\beta}_{5/2}(p_u) \nonumber \\ & & \times \, p_{3,\alpha} p_{3,\beta} \gamma_{5} u_p F(p_u), 
	\\
	\mathcal{M}^s_{Y_{5/2^-}} \!\! &=& -f_I \frac{f_{Y_{5/2^-} N K} f_{Y_{5/2^-} \Xi^* K}}{m^3_\pi} \bar{u}_{\Xi^*,\mu} p_{3,\nu} \gamma_{5} \nonumber \\ & & \times \, G^{\mu\nu\alpha\beta}_{5/2}(p_s) p_{1,\alpha} p_{1,\beta} u_p F(p_s), 
	\\
	\mathcal{M}^u_{Y_{5/2^-}} \!\! &=& f_I \frac{f_{Y_{5/2^-} N K} f_{Y_{5/2^-} \Xi^* K}}{m^3_\pi} \bar{u}_{\Xi^*,\mu} p_{1,\nu} \gamma_{5} \nonumber \\ & & \times \, G^{\mu\nu\alpha\beta}_{5/2}(p_u) p_{3,\alpha} p_{3,\beta} u_p F(p_u), 
\end{eqnarray}
where $f_I$ represents the product of the isospin factors associated with the $YNK$ and $Y\Xi^*K$ vertices. The values of $f_I$ for different final states, intermediate hyperons ($\Lambda^{(*)}$, $\Sigma^{(*)}$), and reaction channels ($s$, $u$) are listed in Table~\ref{tab:ifactor}. A zero entry in this table indicates a forbidden process; for example, $u$-channel $\Lambda^{(*)}$ exchange is prohibited for neutral final states by charge conservation. Furthermore, following the treatment in Refs.~\cite{Sharov:2011xq,Guo:2025ibo}, we treat the product of coupling constants $f_{Y N K}$ and $f_{Y \Xi^* K}$ as a single effective parameter $g_Y$, as these couplings always appear together in any given amplitude.

\begin{table}[htb]
      \caption{Isospin factor products for the different amplitudes.}
      \renewcommand{\arraystretch}{1.2}
      \begin{tabular*}{\columnwidth}{@{\extracolsep\fill}ccccc}
      \hline\hline
      Final States & $\Lambda^{(*)}_s$ & $\Lambda^{(*)}_u$ & $\Sigma^{(*)}_s$ & $\Sigma^{(*)}_u$ \\
      \hline
      charged & $+1$ & $+1$ & $-1$ & $-1$  \\ 
      neutral & $+1$ & $0$  & $+1$ & $-2$  \\ 
      \hline\hline
      \end{tabular*}
      \label{tab:ifactor}
\end{table}

For the triangle-loop diagrams depicted in Fig.~\ref{fig:loop}, we introduce a hypothetical $K^*\Xi$ molecular state $\Lambda^*$ with $J^P=3/2^-$, which is strongly coupled to the $K^*\Xi$ channel via $S$-wave interactions and thereby generates a triangle singularity. The relevant effective Lagrangian densities are given by
\begin{eqnarray}
    \mathcal{L}_{\Lambda^*_{3/2^{-}} \Xi K^*} \!\!\! &=& \!\!\! g_{\Lambda^*_{3 / 2^{-}} \Xi K^*} \bar{\Xi} K^*_\mu \Lambda^{*\mu}_{3 / 2^{-}} + \mathrm{H.c.}, \\
    \mathcal{L}_{\Xi^*\Xi\pi} \!\!\! &=& \!\!\! -\frac{g_{\Xi^*\Xi\pi}}{m_\pi} \bar{\Xi}^*_\mu \vec{\tau} \cdot \partial^\mu \vec{\pi} \Xi + \mathrm{H.c.}, \\
    \mathcal{L}_{K^* K \pi} \!\!\! &=& \!\!\! -g_{K^* K \pi} (\vec{\pi} \cdot \vec{\tau} \partial_\mu \bar{K} - \bar{K} \partial_\mu \vec{\pi} \cdot \vec{\tau}) K^{*\mu}.
\end{eqnarray}
The resulting amplitudes for the triangle-loop diagrams take the form
\begin{eqnarray}
    \mathcal{M}^{\text{Loop}}_{j,k} \!\!\! &=& \!\!\!
    -i f_{j,k} \frac{f_{\Lambda^* N K} g_{\Lambda^* \Xi K^*} g_{K^* K \pi} g_{\Xi^*\Xi\pi}}{m^2_\pi} \bar{u}_{\Xi^*,\mu} \nonumber \\ & & \!\!\! \times \, \int \frac{\mathrm{d}^{4} q}{(2 \pi)^4} q^\mu G_{1/2}(p_\Xi)G_{3/2,\nu\alpha}(p_s) p^\alpha_1 \nonumber \\ & & \!\!\! \times  \, G^{\beta\nu}_{1}(p_{K^*}) \left( p_{3,\beta} - q_\beta \right) \! \gamma_5 u_p F(p_\pi) F(p_s),
\end{eqnarray}
where the subscripts $j,k \in \{c,n\}$ label the charge states of the final-state kaon ($j$) and the intermediate pion ($k$), respectively, with $c$ and $n$ denoting charged and neutral ones. In our calculations, the corresponding isospin factors are $f_{c,n}=f_{n,n}=1$ and $f_{c,c}=f_{n,c}=2$. The coupling constants $g_{K^* K \pi}=3.25$~\cite{Xie:2013wfa} and $g_{\Xi^*\Xi\pi}=0.60$ are extracted from the experimentally measured partial decay widths of $K^* \to K \pi$ and $\Xi^* \to \Xi \pi$, respectively. In analogy to the tree-level treatment, the product $f_{\Lambda^* N K} g_{\Lambda^* \Xi K^*}$ is treated as a single effective parameter $g_{\Lambda^*}$, which is to be determined from fits to the experimental data. Furthermore, we ignore form factors for the internal $K^*$ and $\Xi$ in the triangle loop, as their off-shell effects are negligible in the energy region of interest. 

Based on the amplitudes derived above, we construct the total amplitudes for Model I and Model II. Model I contains solely the tree-level contributions, while Model II comprises both the tree-level and triangle-loop contributions. The cross section is then evaluated via
\begin{eqnarray}
      d\sigma &=& \frac{1}{4} \frac{m_2 m_4}{\sqrt{(p_1 \cdot p_2)^2 -m^2_1 m^2_2}} \frac{1}{(2\pi)^2} \sum_{\lambda_2,\lambda_4} {|\mathcal{M}^{\text{Total}}_{\lambda_2,\lambda_4}|}^2 \nonumber \\ && \times \frac{d^3p_3}{2E_{3}} \frac{d^3p_4}{E_{4}} \delta^4(P_i-P_f),
\end{eqnarray}
where $\mathcal{M}^{\text{Total}}$ is the total amplitude; $P_i$ and $P_f$ denote the total four-momenta of the initial and final states, respectively; and $\lambda_2$ ($\lambda_4$) labels the helicity of the initial (final) baryon.

As discussed in Sec.~\ref{sec:int}, the TS-induced spin effects in the final-state $\Xi^*$ offer a means to discriminate between Model I and Model II, and can thereby provide evidence for the existence of the $K^*\Xi$ molecular state. Here, we evaluate the spin density matrix elements (SDMEs) of the $\Xi^*$ in the center-of-mass (c.m.) frame using the helicity states. The SDME $\rho_{\lambda \lambda'}$ as a function of the c.m. energy $\sqrt{s}$ is defined as
\begin{eqnarray}
      \rho_{\lambda \lambda'}(\sqrt{s}) &=& \frac{\int {\rm d} \Omega_{K} \sum_{\lambda_2} \mathcal{M}^{\text{Total}}_{\lambda_2,\lambda} \mathcal{M}^{\text{Total}*}_{\lambda_2,\lambda'}}{\int{\rm d} \Omega_{K} \sum_{\lambda_2,\lambda_4}|\mathcal{M}^{\text{Total}}_{\lambda_2,\lambda_4}|^2},\label{eq:rho}
\end{eqnarray}
where $\Omega_{K}$ represents the solid angle of the emitted $K$ meson. The diagonal elements $\rho_{11}$ and $\rho_{33}$ correspond to the probability of finding the $\Xi^*$ in the helicity-$1/2$ and helicity-$3/2$ states, respectively. In this work, we focus on the spin observable $P_{\Xi^*}$, defined by
\begin{eqnarray}
      P_{\Xi^*} &=& \frac{\rho_{11}-\rho_{33}}{\rho_{11}+\rho_{33}},\label{eq:Pol}
\end{eqnarray}
which quantifies the asymmetry of the probabilities of the $\Xi^*$ having the helicities $1/2$ and $3/2$, and takes values ranging from $-1$ to $1$. Furthermore, by considering the two-body decay of the $\Xi^*$ in its rest frame, $\rho_{33}$ can be extracted from the angular distribution of the decay products through the following formula~\cite{Jacob:1959at,Schilling:1969um,Thomas:1973uh,Kim:2017hhm}
\begin{equation}
    W(\cos\theta) = \frac{1}{4} \left[ \left( 1+4\rho_{33} \right) + \left( 3-12\rho_{33} \right) \cos^2\theta \right],\label{eq:AD}
\end{equation}
while $\rho_{11}$ is subsequently fixed by the normalization relation $\rho_{11} + \rho_{33} = 1/2$.

\begin{figure}[tbp]
      \includegraphics[width=0.8\columnwidth]{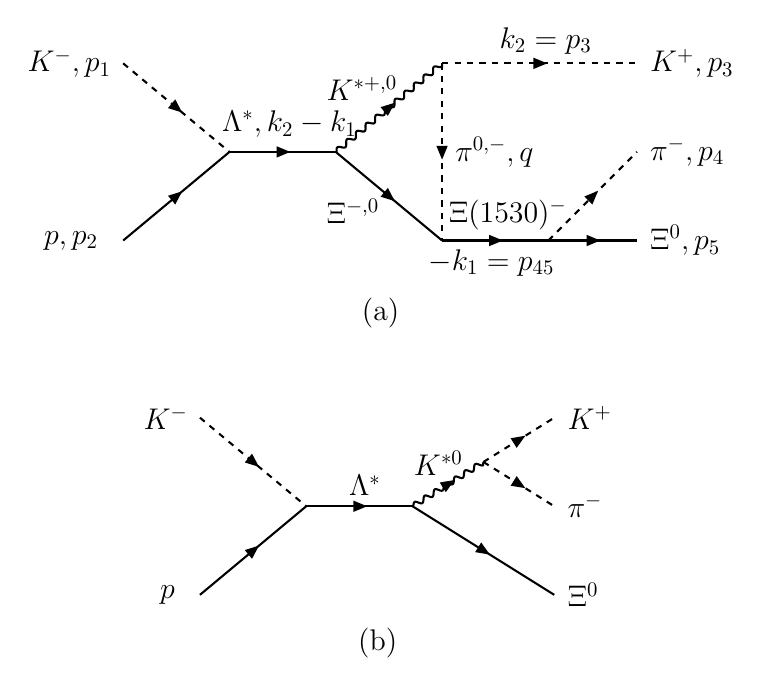}
	\caption{(a) Triangle-loop and (b) tree-level Feynman diagrams for the reaction $K^- p \to K^+ \pi^- \Xi^0$.}
	\label{fig:loop3}
\end{figure}

To extract $\rho_{33}$, we consider the specific reaction $K^- p \to K^+ \pi^- \Xi^0$ and determine the angular distribution of the $\pi^-$ in the $\Xi^0 \pi^-$ rest frame. As illustrated in Fig.~\ref{fig:loop3}(a), the triangle-loop involves $\pi^0 \Xi^- \to \pi^- \Xi^0$ and $\pi^- \Xi^0 \to \pi^- \Xi^0$ subprocesses. According to Schmid's theorem~\cite{Schmid:1967ojm}, the triangle-loop contribution including the elastic subprocess $\pi^- \Xi^0 \to \pi^- \Xi^0$ is negligible compared with the corresponding tree-level diagram shown in Fig.~\ref{fig:loop3}(b). Consequently, the tree-level background contribution may obscure the characteristic spin features at the TS position. Although Ref.~\cite{Debastiani:2018xoi} demonstrated that Schmid's theorem holds strictly only in the limit $\Gamma_{K^*} \to 0$ and that the background contributions can be suppressed by applying an appropriate cut on the invariant mass $M_{\Xi\pi}$, it is still necessary to discuss the impact of this background on the extraction of spin observables. Using the Feynman diagrams, effective Lagrangian densities, and two-body amplitudes introduced above, we construct the complete three-body amplitudes, from which the differential cross sections are then evaluated via~\cite{Wang:2020rsb,Wang:2023lnb}
\begin{eqnarray}
      d\sigma &=& \frac{1}{4} \frac{m_2 m_5}{\sqrt{(p_1 \cdot p_2)^2 -m^2_1 m^2_2}} \frac{1}{(2\pi)^5} \sum_{\lambda_2,\lambda_5} {|\mathcal{M}^{\text{Total}}_{\lambda_2,\lambda_5}|}^2 \nonumber \\ && \times \, \frac{d^3p_3}{2E_3} \frac{d^3p_4}{2E_4} \frac{d^3p_5}{E_5} \delta^4(P_i-P_f).
\end{eqnarray}

\section{Results and discussions}\label{sec:res}

In this section, we present the cross-section fitting results for the resonance and TS models to elucidate the production mechanism of the observed structure near $\sqrt{s}=2.25$~GeV, and show the roles of possible $K\Xi^*$ and $K^*\Xi$ molecular states. We then predict the energy dependence of the spin observable $P_{\Xi^*}$ and discuss the prospects for distinguishing between different models using this quantity. Finally, we demonstrate how spin observables can be extracted from angular distributions in three-body final states and evaluate the impact of background contributions on the final results.  

For Model I, the observed structures in the cross sections are assumed to originate solely from tree-level resonance production. To reproduce the experimental data with as few states as possible, we include contributions from the $\Lambda(1116)$, $\Lambda(2050)$, $\Lambda(2110)$, $\Lambda(2325)$, $\Sigma(1193)$, and $\Sigma(2250)$ resonances. The total amplitudes for Model I then read 
\begin{eqnarray}\label{amp:R}
    \mathcal{M}^{\text{Total}}_{\text{I},j} \!\! &=& 
    \!\! \mathcal{M}_{\Lambda(1116)} + \mathcal{M}_{\Lambda(2110)} + \mathcal{M}_{\Lambda(2325)} + \mathcal{M}_{\Sigma(1193)} \nonumber \\ & & \!\! + \, \mathcal{M}_{\Sigma(2250)} + e^{i\phi_j} \mathcal{M}_{\Lambda(2050)}.
\end{eqnarray}
Because the $\Lambda(2050)3/2^-$ is treated as the $K\Xi^*$ molecular state, a relative phase angle $\phi_j$ is introduced to effectively account for possible loop effects, where $j=c$ ($n$) labels the charged (neutral) final states. In this fit, the six products of coupling constants $g_Y = f_{YNK} f_{Y\Xi^*K}$ are treated as free parameters and determined from the cross-section data; their best-fit values are listed in Table~\ref{tab:Parameter:R}. 

\begin{figure*}[tbp]
      \includegraphics[width=0.9\textwidth]{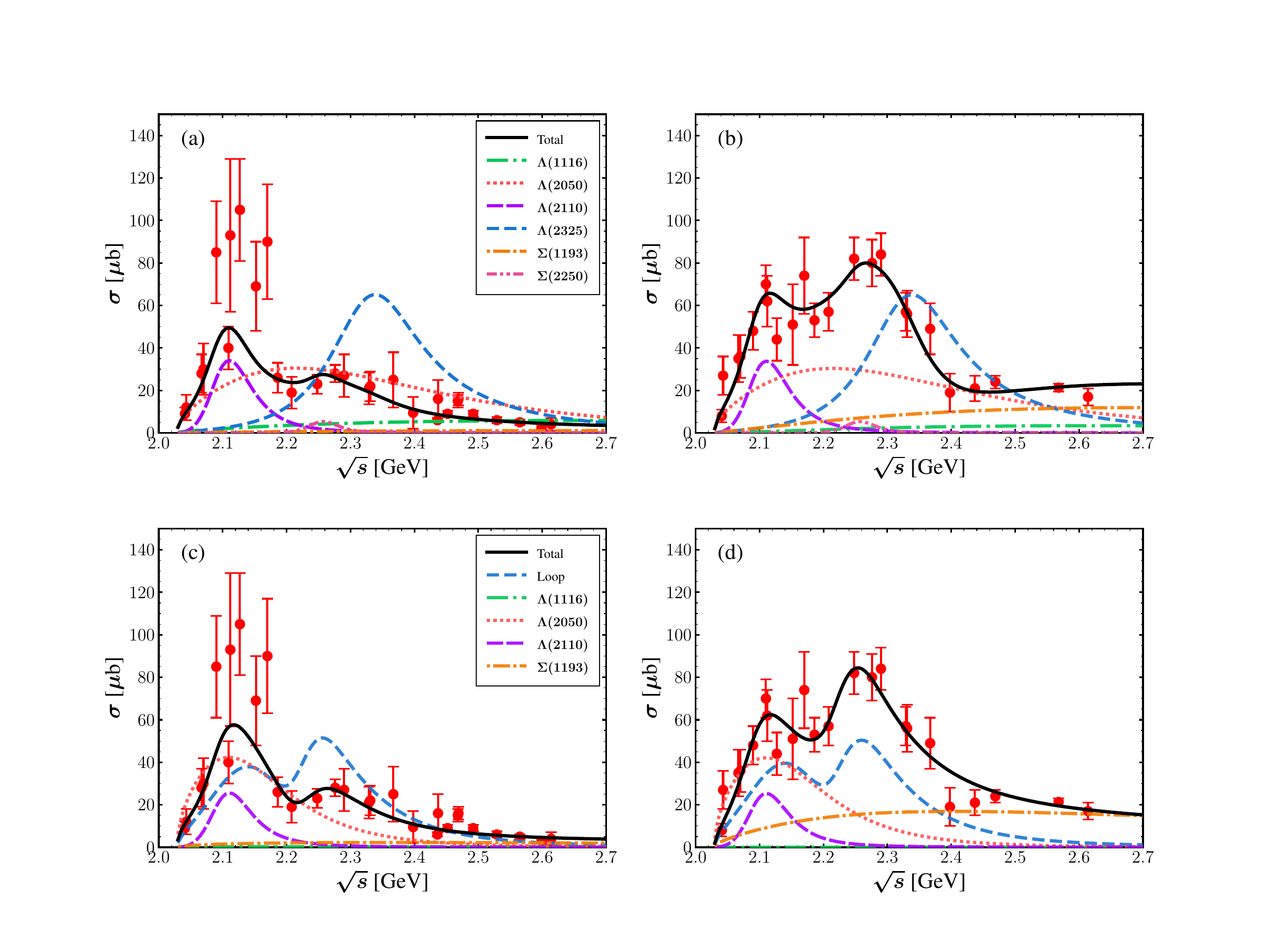}
	\caption{Total cross sections for $K^- p \to K \Xi(1530)$ as a function of c.m. energy $\sqrt{s}$. Panels (a) and (b) show Model I (resonance model) results for $K^+ \Xi(1530)^-$ and $K^0 \Xi(1530)^0$, respectively; panels (c) and (d) show the corresponding Model II (TS model) results. Experimental data are taken from Refs.~\cite{Berge:1966zz,Dauber:1969hg,Briefel:1977bp}.}
	\label{fig:CS}
\end{figure*}

The final fitting results are presented in Fig.~\ref{fig:CS}. As shown in Fig.~\ref{fig:CS}(a) and~(b), the peaks at $2.1$~GeV and $2.25$~GeV originate from the $\Lambda(2110)$ and $\Sigma(2250)$ resonances, respectively. The rapid growth near threshold is attributed to the $\Lambda(2050)$, in agreement with the $K\Xi^*$ molecular-state hypothesis. The ground-state hyperons $\Lambda$ and $\Sigma$ play essential roles in the $K^- p \to K^+ \Xi(1530)^-$ and $K^- p \to K^0 \Xi(1530)^0$ reactions, respectively, in the high-energy region. Notably, interference effects between the $\Lambda(2050)$ and $\Lambda(2325)$ resonances are responsible for the significant differences between the two charge channels; for instance, the $\Lambda(2325)$ structure manifests as a small bump in Fig.~\ref{fig:CS}(a) but as a steep drop in Fig.~\ref{fig:CS}(b). Overall, the tree-level resonance model satisfactorily reproduces the available data, consistent with the conclusions of Ref.~\cite{Guo:2025ibo}.

\begin{table}[htb]
      \caption{Fitted parameters for Model I.}
      \renewcommand{\arraystretch}{1.2}
      \begin{tabular*}{\columnwidth}{@{\extracolsep\fill}cccc}
      \hline\hline
      Parameter & Value & Parameter & Value \\
      \hline
      $g_{\Lambda(1116)}$ & $4.71\times 10^{-2}$ & $g_{\Sigma(2250)}$ & $1.10\times 10^{-2}$  \\ 
      $g_{\Lambda(2050)}$ & $4.30\times 10^{-1}$ & $\phi_c$ & $2.90$  \\ 
      $g_{\Lambda(2110)}$ & $-5.46\times 10^{-2}$ & $\phi_n$ & $3.12$  \\
      $g_{\Lambda(2325)}$ & $1.33\times 10^{-1}$ & $\Lambda_B$~[GeV] & $1.8$  \\
      $g_{\Sigma(1193)}$ & $2.20\times 10^{-2}$ & $\chi^2/\text{dof}$ & $1.60$ \\ 
      \hline\hline
      \end{tabular*}
      \label{tab:Parameter:R}
\end{table}

For Model II, we assume that the peak near $\sqrt{s}=2.25$~GeV arises from the TS mechanism illustrated in Fig.~\ref{fig:loop}, rather than from the $\Sigma(2250)$ resonance. This mechanism requires an intermediate state with strong coupling to the $K^*\Xi$ channel. Based on the findings of Ref.~\cite{Dong:2021bvy}, in addition to the $K\Xi^*$ system, the $K^*\Xi$ can also form a molecular state with quantum numbers $I (J^P) = 0 (1/2^-)$ or $0 (3/2^-)$. Accordingly, we introduce a putative $\Lambda^*$ molecular state to trigger this TS process. In what follows, we present results only for the $J^P=3/2^-$ assignment, as the fit strongly favors this case. To minimize tree-level contributions, we retain only the $\Lambda(1116)$, $\Lambda(2050)$, $\Lambda(2110)$, and $\Sigma(1193)$ as intermediate particles. The total amplitudes for Model II read
\begin{eqnarray}\label{amp:TS}
    \mathcal{M}^{\text{Total}}_{\text{II},j} &=& 
    \mathcal{M}_{\Lambda(1116)} + \mathcal{M}_{\Lambda(2110)} + \mathcal{M}_{\Sigma(1193)} \nonumber \\ & & + \, \mathcal{M}^{\text{Loop}} + e^{i\phi_j} \mathcal{M}_{\Lambda(2050)}.
\end{eqnarray}
The free parameters characterizing the putative $K^*\Xi$ molecular state $\Lambda^*$---its mass $m_{\Lambda^*}$, width $\Gamma_{\Lambda^*}$, and the coupling product $g_{\Lambda^*}=f_{\Lambda^*NK} g_{\Lambda^*\Xi K^*}$---are determined by fitting to the experimental data; the best-fit values are listed in Table~\ref{tab:Parameter:TS}. 

\begin{table}[htb]
      \caption{Fitted parameters for Model II.}
      \renewcommand{\arraystretch}{1.2}
      \begin{tabular*}{\columnwidth}{@{\extracolsep\fill}cccc}
      \hline\hline
      Parameter & Value & Parameter & Value \\
      \hline
      $g_{\Lambda(1116)}$ & $-4.88\times 10^{-2}$ & $m_{\Lambda^*}$~[MeV] & $2151.4$ \\ 
      $g_{\Lambda(2050)}$ & $6.55\times 10^{-1}$ & $\Gamma_{\Lambda^*}$~[MeV] & $306.94$ \\ 
      $g_{\Lambda(2110)}$ & $-4.73\times 10^{-2}$ & $g_{\Lambda^*}$ & $-0.79$  \\
      $g_{\Sigma(1193)}$ & $1.83\times 10^{-1}$ & $\Lambda_\pi$~[GeV] & $2.57$  \\
      $\phi_c$ & $0.15$ & $\Lambda_B$~[GeV] & $0.91$  \\
      $\phi_n$ & $1.80$ & $\chi^2/\text{dof}$ & $1.58$  \\
      \hline\hline
      \end{tabular*}
      \label{tab:Parameter:TS}
\end{table}

As shown by the blue dashed lines in Fig.~\ref{fig:CS}(c) and~(d), the triangle-loop process dominates the energy region from $2.2$ to $2.4$~GeV. Therefore, the peak near $2.25$~GeV is attributed entirely to the TS mechanism. In contrast, the structure near $2.1$~GeV arises from three sources: the $\Lambda(2050)$, $\Lambda(2110)$, and the putative $\Lambda^*$ molecular state. Hereafter, we denote this state as $\Lambda(2150)$, in accordance with its fitted mass of $2151.4$~MeV. The poor description of this structure in the $K^+\Xi(1530)^-$ channel is mainly due to the large uncertainties in the experimental data for $2.09 \le \sqrt{s} \le 2.17$~GeV, in particular the inconsistency between the data point at $2.11$~GeV and its neighbors. This description could be improved by removing the data at $\sqrt{s}=2.11$~GeV, as done in Ref.~\cite{Guo:2025ibo}, and the peak could then be reproduced by enhancing the $\Lambda(2110)$ contribution. However, we do not adopt such a treatment in this work because the uncertainties tied to the $\Lambda(2110)$ parameters do not affect our main conclusions regarding the $K^*\Xi$ molecular state and the TS peak. Furthermore, although the ground-state $\Lambda$ and $\Sigma$ contributions are small for the charged final states, the $\Sigma$ contribution is non-negligible for the neutral final states in the high-energy region. Neither model can fully account for the structures observed near $2.4$~GeV, namely a dip in the $K^+\Xi(1530)^-$ channel and a sharp drop in the $K^0\Xi(1530)^0$ channel. Since the $K^*\Xi(1530)$ threshold is located at $2.43$~GeV, we conjecture that a molecular state or threshold effect may be responsible for these features. This issue, however, lies beyond the scope of the present study owing to the limited experimental data and will be addressed in future work. 

\begin{figure*}[tbp]
      \includegraphics[width=0.9\textwidth]{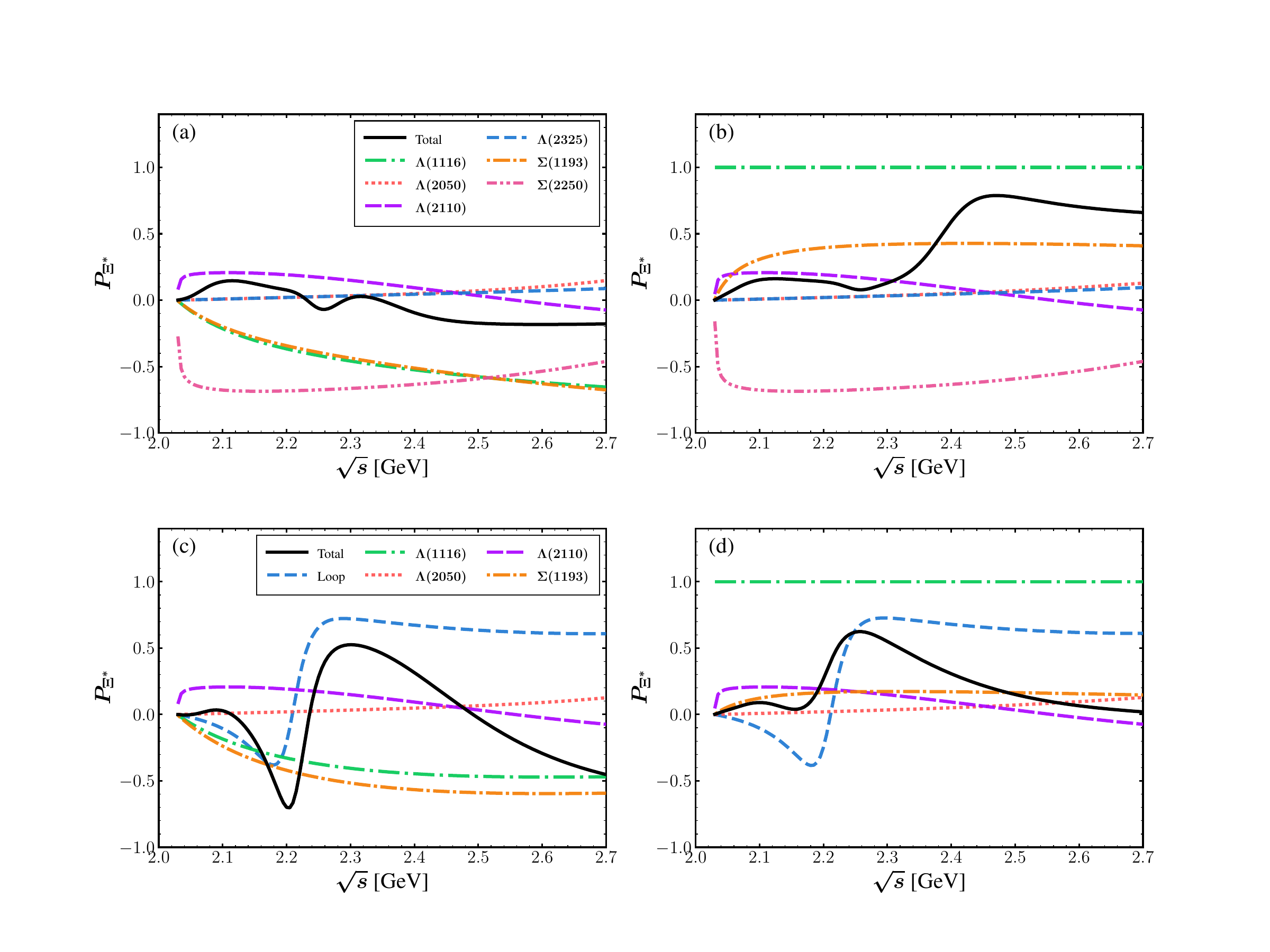}
	\caption{Spin observable $P_{\Xi^*}$ as a function of $\sqrt{s}$ for the different models and final states. The panel layout follows that of Fig.~\ref{fig:CS}. }
	\label{fig:Pol}
\end{figure*}

A comparison of the reduced $\chi^2$ values reveals that, although Model II yields a better fit, both models can reasonably describe the experimental data given the large uncertainties. Consequently, the cross-section data alone are insufficient to discriminate between the two models. Fortunately, the TS mechanism generates a distinct spin effect in $\Xi^*$ production that can distinguish between the resonance and TS models~\cite{Wang:2022wdm,Wang:2023xua}. According to the Coleman-Norton theorem~\cite{Coleman:1965xm}, a TS occurs only when the triangle-loop process depicted in Fig.~\ref{fig:loop} proceeds in a classical manner. Hence, at the TS, the intermediate $\pi$ and $\Xi$ are collinear with the final $\Xi^*$. This kinematic condition restricts the helicity of $\Xi^*$ to $\pm 1/2$ only, via angular momentum conservation along the momentum direction of the $\Xi^*$ in the c.m. frame. In contrast, when these kinematic conditions are not satisfied, i.e., away from the TS position, the helicity of $\Xi^*$ is no longer constrained to $\pm 1/2$. Based on this analysis, the spin asymmetry observable $P_{\Xi^*}$ associated with the loop amplitude exhibits strong energy dependence and approaches unity at the TS. 

In Fig.~\ref{fig:Pol}, we present the spin asymmetry $P_{\Xi^*}$ as a function of $\sqrt{s}$ for both models. For each individual tree-level contribution, the corresponding $P_{\Xi^*}$ curve exhibits weak energy dependence, varying significantly only in a narrow region near threshold. Consequently, the energy dependence of the total amplitude in Model I arises primarily from the varying relative strengths of the different resonances. By contrast, the loop-amplitude contribution displays pronounced intrinsic energy dependence: starting from threshold, the value of $P_{\Xi^*}$ decreases gradually with increasing energy, whereas the TS mechanism drives it toward a large value\footnote{In practice, $P_{\Xi^*}$ does not reach unity exactly at the TS for two reasons: (i) the finite width of the intermediate $K^*$ shifts the TS position away from the real axis, and (ii) the non-negligible off-shell effect of the exchanged $\pi$ meson (with cutoff $\Lambda_\pi = 2.57$~GeV) reduces the relative TS contribution.} near $\sqrt{s}=2.25$~GeV, producing a sharp rise between $2.2$ and $2.3$~GeV. Because the loop contribution dominates the reaction in Model II, the energy dependence of the total amplitude primarily reflects the TS mechanism. Moreover, interference between the triangle-loop and tree-level amplitudes enhances this rapidly growing structure in the $K^+ \Xi(1530)^-$ channel while suppressing it in the $K^0 \Xi(1530)^0$ channel. This pronounced structure thus provides a clear signature for discriminating between the two models.

\begin{figure}[tbp]
      \includegraphics[width=0.9\columnwidth]{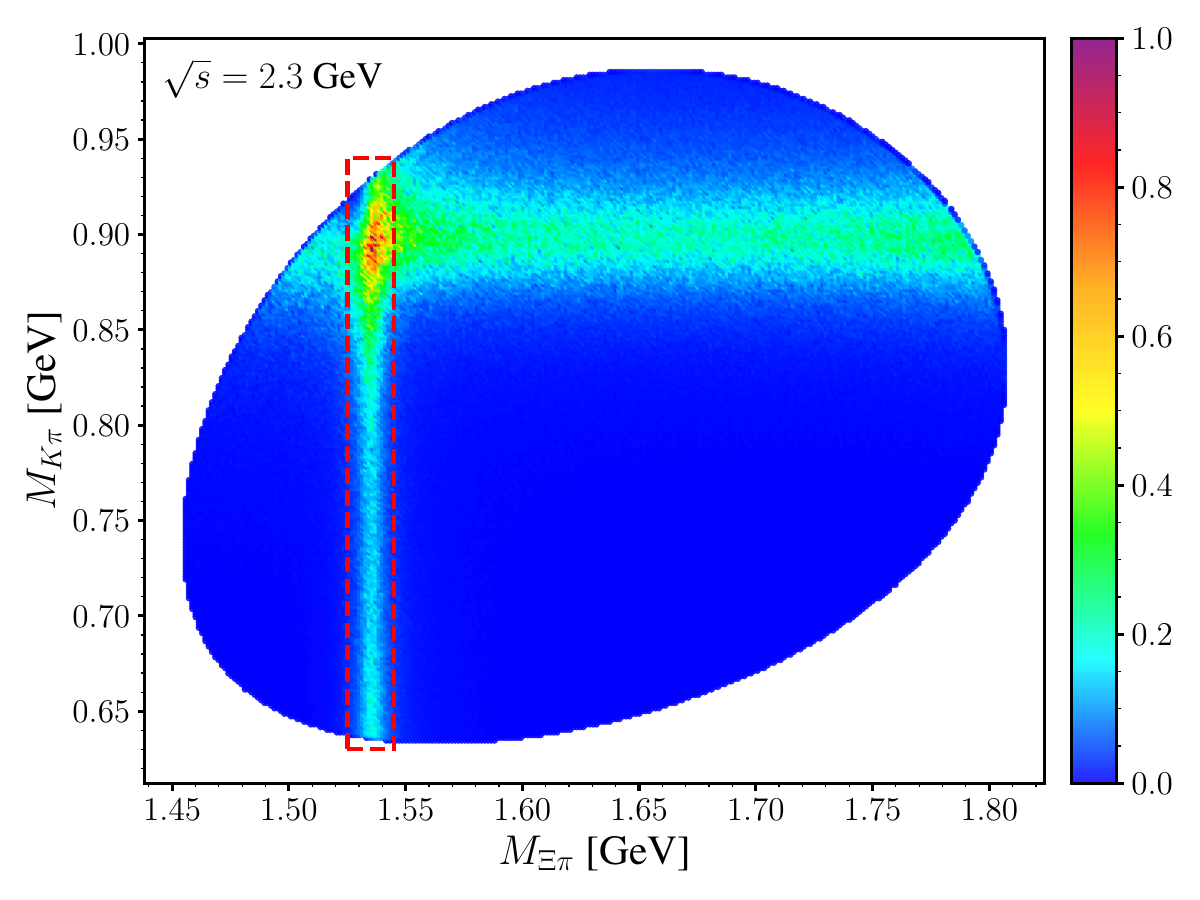}
	\caption{Dalitz plot for $K^- p \to K^+ \pi^- \Xi^0$ at $\sqrt{s}=2.3$~GeV in Model II. The red dashed box indicates the kinematic cut on the $\Xi \pi$ invariant mass, selecting events within the $\Xi^*$ resonance region ($m_{\Xi^*} - \Gamma_{\Xi^*} < M_{\Xi \pi} < m_{\Xi^*} + \Gamma_{\Xi^*}$) used to reconstruct the pion angular distribution in the $\Xi \pi$ rest frame.}
	\label{fig:Dalitz}
\end{figure}

We identify the spin asymmetry $P_{\Xi^*}$ as a key observable for discriminating between the resonance and TS models, thereby offering crucial insight into the production mechanism of the structure near $\sqrt{s}=2.25$~GeV. Experimentally, the $\Xi^*$ spin information is extracted from the angular distribution of its decay products in the $\Xi^*$ rest frame via Eq.~\eqref{eq:AD}. To facilitate comparison with future measurements, we incorporate the $\Xi^*$ decay into our calculation and assess the impact of background contributions in the three-body final states. As an illustration, we calculate the $K^+ \pi^- \Xi^0$ production process within Model II, including both the tree-level and triangle-loop diagrams shown in Fig.~\ref{fig:loop3} (additional tree-level diagrams corresponding to Fig.~\ref{fig:tree} are included in the calculation but omitted from the figure for clarity). The tree-level process involving the intermediate $K^*$ meson generates significant background that may obscure the TS signature in the angular distribution. To suppress this background, we analyze the Dalitz plot at $\sqrt{s}=2.3$~GeV shown in Fig.~\ref{fig:Dalitz}, which exhibits distinct, well-separated band structures corresponding to the $\Xi^*$ and $K^*$ intermediate states. By applying the kinematic cut $m_{\Xi^*} - \Gamma_{\Xi^*} < M_{\Xi \pi} < m_{\Xi^*} + \Gamma_{\Xi^*}$ (indicated by the red dashed box) to select $\Xi^*$ events while excluding $K^*$ contributions, we reconstruct the angular distribution in the $\Xi\pi$ rest frame. This procedure preserves the characteristic TS features in the extracted spin observables.

\begin{figure}[tbp]
      \includegraphics[width=0.9\columnwidth]{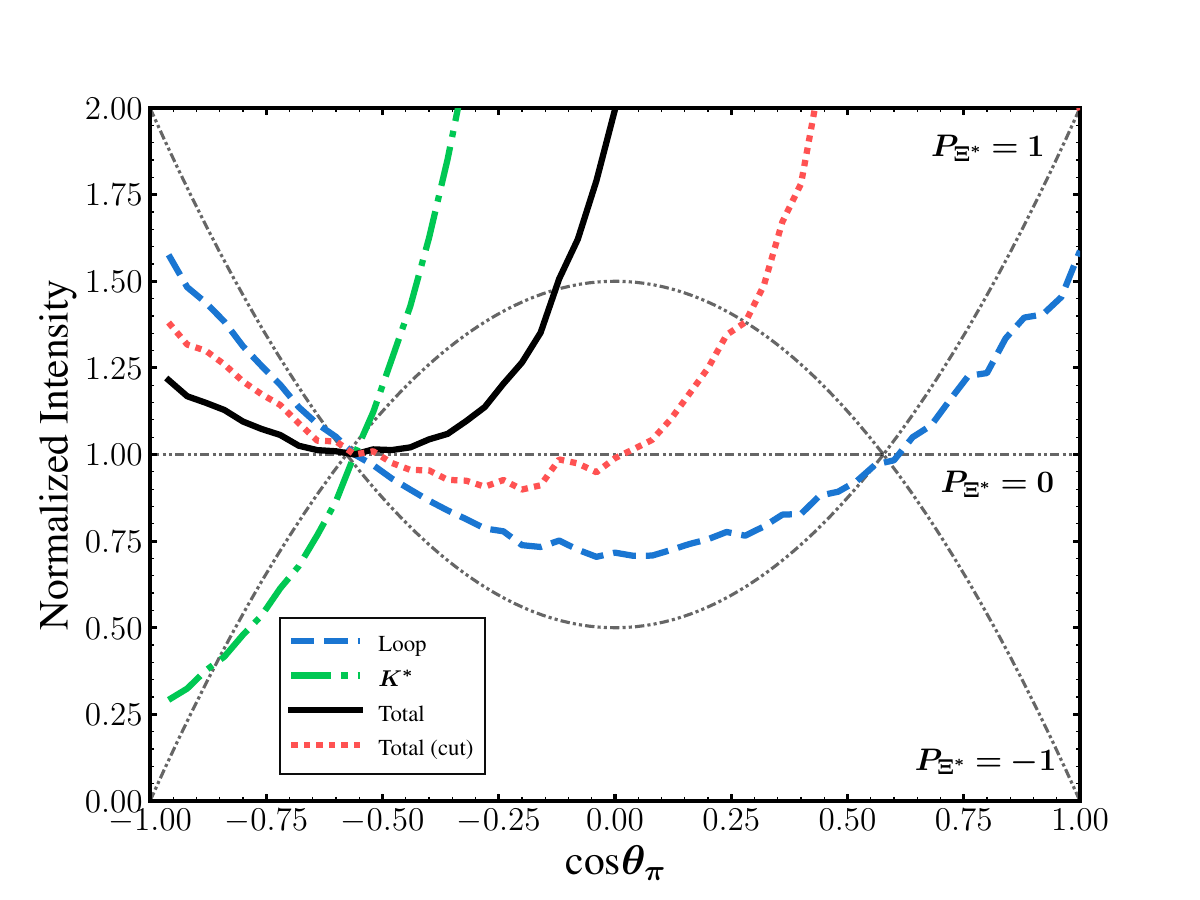}
	\caption{Angular distribution of the pion in the $\Xi \pi$ rest frame for Model II at $\sqrt{s}=2.3$~GeV. $\theta_\pi$ is the angle between the pion momentum and the direction opposite to the kaon momentum. }
	\label{fig:AD:TS}
\end{figure}

To quantify the improvement achieved by the kinematic cut, we compare the angular distributions obtained with and without the cut. According to Eq.~\eqref{eq:AD}, the angular distribution is independent of the spin density matrix element $\rho_{33}$ at $\cos\theta=\pm 1/\sqrt{3}$. Therefore, all angular distributions are normalized to unity at $\cos\theta_\pi=-1/\sqrt{3}$, as shown in Fig.~\ref{fig:AD:TS}. The three black dashed lines indicate the expectations for $P_{\Xi^*}=+1$, $0$, and $-1$ (from top to bottom at $\cos\theta_\pi=1$). The blue dashed and green dash-dotted curves represent the triangle-loop and $K^*$-mediated background amplitudes, respectively. The triangle-loop amplitude yields a symmetric distribution with $P_{\Xi^*}$ close to unity, consistent with the predictions in Fig.~\ref{fig:Pol}. By contrast, the background exhibits an asymmetric distribution strongly enhanced in the forward direction ($\cos\theta_\pi \to 1$). When the angular distribution is reconstructed using all events from the Dalitz plot, the substantial background contributions distort the final result (black solid line) away from the theoretical form of Eq.~\eqref{eq:AD}, precluding reliable extraction of the $\Xi^*$ spin information. After applying the kinematic cut, the individual angular distributions retain nearly the same shapes; hence, only the total amplitude result (red dotted line) is presented. The cut substantially reduces the relative background contribution. Although a non-negligible background persists in the forward region, the triangle-loop amplitude dominates the backward region ($\cos\theta_\pi \to -1$), enabling the extraction of the correct spin observables.

\begin{figure}[tbp]
      \includegraphics[width=0.9\columnwidth]{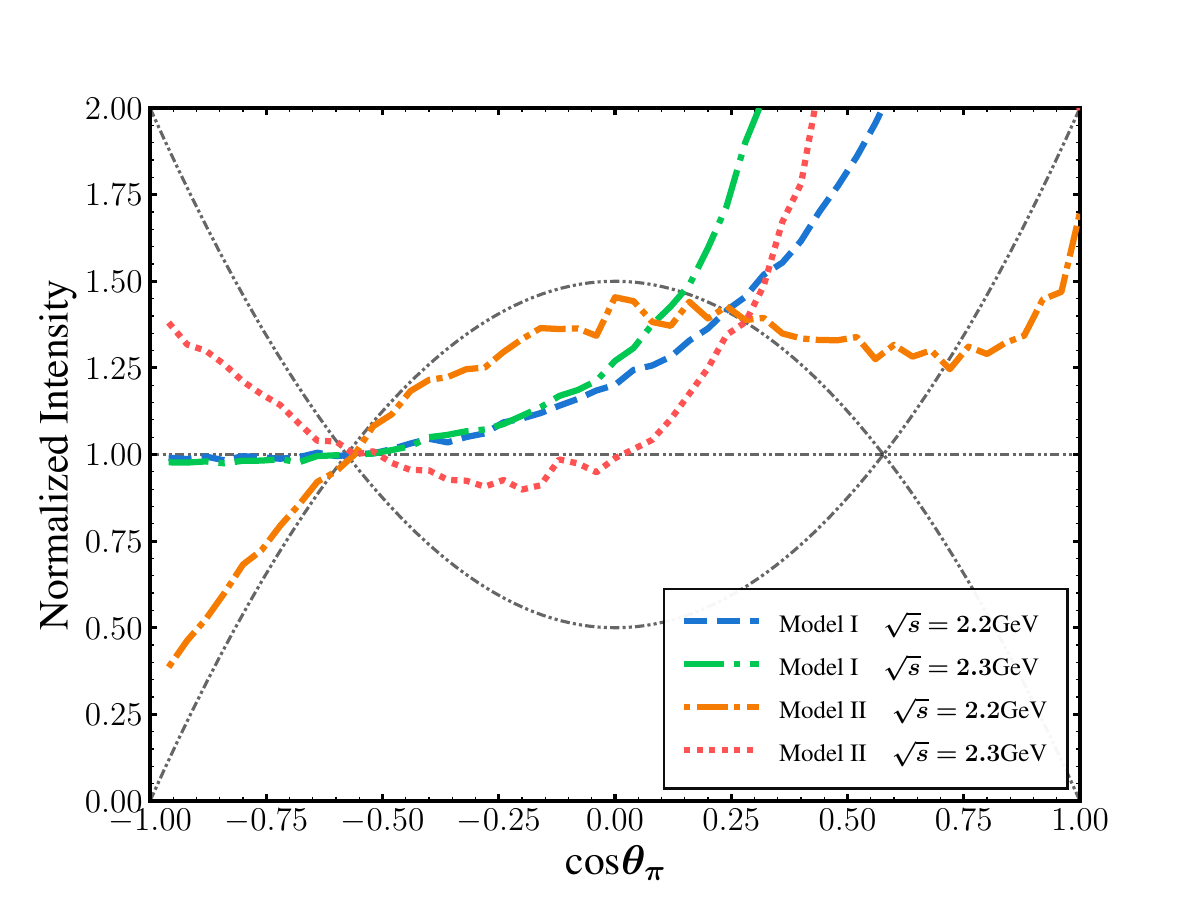}
	\caption{Angular distributions of the pion in the $\Xi\pi$ rest frame for both models at selected c.m. energies, obtained with the kinematic cut on the $\Xi\pi$ invariant mass.}
	\label{fig:AD:All}
\end{figure}

Based on the above analysis, $P_{\Xi^*}$ can be experimentally determined from the shape of the angular distribution in the backward region. Furthermore, its variation between $\sqrt{s}=2.2$ and $\sqrt{s}=2.3$~GeV provides a clear signature for distinguishing between the resonance and TS models. To demonstrate this, we apply the same kinematic cut to the three-body final states in both models and extract $P_{\Xi^*}$ at these two energies. However, the fit in Model I does not constrain the strength of the background contribution shown in Fig.~\ref{fig:loop3}(b). We therefore estimate this background for Model I by adopting the corresponding contribution from Model II. Although this procedure may overestimate the background in Model I, our conclusions remain unaffected. The resulting angular distributions are shown in Fig.~\ref{fig:AD:All}, where the extracted values in the backward region are consistent with the predicted $P_{\Xi^*}$. For Model I, $P_{\Xi^*}$ remains near zero at both energies, whereas for Model II, it approaches $-1$ at $\sqrt{s}=2.2$~GeV and $+1$ at $\sqrt{s}=2.3$~GeV. Overall, the pronounced differences in the angular distributions between the two models provide a definitive means to determine whether the peak near $\sqrt{s}=2.25$~GeV in the $K^- p \to K \Xi(1530)$ reaction arises from resonance production or the TS mechanism.

\section{Summary and conclusion}  \label{sec:sum}

In this work, we have investigated possible $K\Xi^*$ and $K^*\Xi$ molecular states, as well as the triangle singularity, in the $K^- p \to K \Xi(1530)$ reaction within an effective Lagrangian approach. We find that the $\Lambda(2050)3/2^-$ can be regarded as the $K\Xi^*$ molecular state and plays an essential role in reproducing the near-threshold enhancement. Furthermore, a $K^*\Xi$ molecular state $\Lambda(2150)$ with $I(J^P)=0(3/2^-)$ can generate a TS through a triangle-loop diagram involving intermediate $K^*$, $\Xi$, and $\pi$ states. Hence, the peak observed in the cross section near $\sqrt{s}=2.25$~GeV can be interpreted either as the $\Sigma(2250)$ resonance or as a TS structure associated with the $\Lambda(2150)$. Although both scenarios provide comparable descriptions of the cross-section data, they can be discriminated by the spin observable $P_{\Xi^*}$. Specifically, the unique kinematic conditions of the TS impose stringent constraints on the $\Xi^*$ helicity, leading $P_{\Xi^*}$ to exhibit a characteristic energy dependence, approaching unity at the TS position, that is absent in the resonance model. The predicted dramatic variation of $P_{\Xi^*}$ between $\sqrt{s}=2.2$ and $2.3$~GeV thus serves as a decisive discriminator between the two mechanisms. We further propose a practical method to extract $P_{\Xi^*}$ from the pion angular distribution in the $\Xi\pi$ rest frame through the study of $K^- p \to K^+ \pi^- \Xi^0$, carefully examining the impact of background contributions. Although the background dominates in certain kinematic regions, an appropriate kinematic cut on the $\Xi\pi$ invariant mass effectively suppresses these contributions, enabling reliable extraction of the spin observables. We encourage future experimental measurements of these spin observables at facilities such as J-PARC, where high-precision data will clarify the dynamical origin of the $2.25$~GeV peak and test the existence of the $K^*\Xi$ molecular state.

\bigskip

\begin{acknowledgments}
We thank Jia-Jun Wu and Feng-Kun Guo for valuable discussions and constructive suggestions. This work is supported by the National Natural Science Foundation of China under Grants No.~12575093 and No.~12175240.
\end{acknowledgments}

\bibliographystyle{apsrev4-1}
\bibliography{Xi(1530).bib}

\end{document}